\documentclass[twocolumn,astrosymb]{aastex631}

\usepackage[T1]{fontenc}
\usepackage{amsmath}

\shorttitle{Dust and Gas Properties from Ringed Disks}
\shortauthors{Eve J.~Lee}

\begin{document}

\title{Probing Dust and Gas Properties Using Ringed Disks}

\correspondingauthor{Eve J.~Lee}
\email{eve.lee@mcgill.ca}

\author[0000-0002-1228-9820]{Eve J. Lee}
\affiliation{Department of Physics and Trottier Space Institute, McGill University, 3600 rue University, H3A 2T8 Montreal QC, Canada}
\affiliation{Trottier Institute for Research on Exoplanets (iREx), Universit\'e de Montr\'eal, Canada}

\begin{abstract}

How rapidly a planet grows in mass and how far they may park from the host star depend sensitively on two non-dimensional parameters: Stokes number St and turbulent $\alpha$. Yet, these parameters remain highly uncertain being difficult or impossible to measure directly. Here, we demonstrate how the ringed disks can be leveraged to obtain St and $\alpha$ separately by constructing a simple toy model that combines dust radial equation of motion under aerodynamic drag and coupling to gas motion with the measured distribution of dust masses in Class 0/I disks. Focusing on known systems with well-resolved dust rings, we find that the range of St and $\alpha$ that are consistent with the measured properties of the rings are small: $10^{-4} \lesssim {\rm St} \lesssim 10^{-2}$ and $10^{-5} \lesssim \alpha \lesssim 10^{-3}$. These low St and $\alpha$ ensure the observed rings are stable against clumping. Even in one marginal case where the formation of bound clumps is possible, further mass growth by pebble accretion is inhibited. Furthermore, the derived low $\alpha$ is consistent with the nearly inviscid regime where Type I migration can be prematurely halted. Our analysis predicts minimal planet population beyond $\sim$10s of au where we observe dust rings and significantly more vigorous planet formation inside $\sim$10 AU, consistent with current exo-giant statistics. We close with discussions on the implications of our results on small planet statistics at large orbital distances.

\end{abstract}

\section{Introduction} \label{sec:intro}

One of the biggest uncertainties in planet formation is their initial conditions. Whether a planet becomes a gas giant or remain a gas-poor sub-Neptune or a rocky super-Earth/Earth is sensitively determined by how rapidly the underlying rocky core can grow \citep[e.g.,][]{Lee19,Lee22-accr}. The initial growth of the planetary cores is expected to be governed by pebble accretion \citep[e.g.,][]{Ormel10,Lambrechts12,Ormel17} whereby particles of Stokes number St $< 1$ settle onto an accreting body via aerodynamic drag. The growth rate by pebble accretion is then necessarily sensitively determined by St, with smaller St slowing down the growth of solid cores. In fact, under pebble accretion and the radial drift of solids in protoplanetary disks \citep[e.g.,][]{Weidenschilling77}, \citet{Lin18} report that a necessary (but not sufficient) condition for the rapid coagulation of massive cores that can generate a gas giant is for the particle Stokes number to be larger than the Shakura-Sunyaev turbulent parameter $\alpha$.

Large $\alpha$ (stronger diffusive turbulence) can puff up the dust disk diluting the solid density and therefore slowing down pebble accretion. More importantly, once $\alpha \lesssim 10^{-4}$--$10^{-3}$, disk-induced Type I migration is expected to stall prematurely as the perturbed gas interior to the planet's orbit continue to pile up before it is diffused, driving the net torque on the planet to zero \citep[e.g.,][]{Rafikov02,Li09,Yu10,Fung17,Fung18}. It follows that the two non-dimensional parameters St and $\alpha$ critically determine nearly all aspects of the early stages of planet formation, from the mass growth of rocky cores to the emplacement of the planet, which altogether determine the final observed masses, radii, and orbital periods of exoplanets.

Yet, direct measurements of St and $\alpha$ are either impossible or challenging. Under the Epstein drag regime, Stokes number is defined as
\begin{equation}
    {\rm St} = \Omega\frac{\rho_s s}{\rho_g c_s}
    \label{eq:St}
\end{equation}
where $\Omega \equiv \sqrt{GM_\star /a^3}$ is the Keplerian orbital frequency, $G$ is the gravitational constant, $M_\star$ is the mass of the host star, $a$ is the orbital distance, $\rho_s$ is the material density of the grain, $s$ is the size of the grain, $\rho_g$ is the volumetric mass density of disk gas, $c_s \equiv \sqrt{kT/\mu m_H}$ is the sound speed, $k$ is the Boltzmann constant, $T$ is the midplane temperature, $\mu$ is the gas mean molecular weight, and $m_H$ is the Hydrogen atomic mass. The wavelength of observation that probes the dust thermal emission can be used as a proxy for the grain size, albeit not exact. At large enough distances (beyond $\sim$10 au), the disk midplane temperature can be approximated to be from heating by stellar irradiation \citep[e.g.,][]{Chiang97,DAlessio98}. Gas measurements are limited and so $\rho_g$ remains largely unknown or poorly constrained, and even when CO measurements can be made \citep[e.g.,][]{Zhang21}, conversion to the total gas mass depends on the uncertain CO-H$_2$ factor.

Measurement of $\alpha$ is equally challenging for protoplanetary disks given that the turbulent dynamics is expected to be highly subsonic. Nevertheless, reports of $\alpha$ based on the depth of the gap in dust emission \citep{Pinte16} and non-thermal broadening of molecular lines \citep{Flaherty15,Flaherty17} find $\alpha \lesssim 10^{-3}$ over a wide range of vertical depth and radial extent. 

Gas kinematic data of sufficiently high resolution are expensive and often require complex chemical modelling to disentagle. In this paper, we seek to derive St and $\alpha$ with simpler and readily available dust emission measurements of ringed disks. Ring-like substructures are common in radio interferometric images of protoplanetary disks \citep[e.g.,][]{ALMA2015,Andrews18}. Modeling the gas pressure perturbation as a Gaussian of width $w$, the observed dust ring can be interpreted as the equilibrium between the aerodynamic drift that collects the particles at the center of the pressure bump and the turbulent diffusion so that the width of the dust ring is \citep{Dullemond18}
\begin{equation}
    w_{\rm d} = w\left[1+\left(\frac{{\rm Sc}{\rm St}}{\alpha}\right)\right]^{-1/2}
    \label{eq:dust-ring-width}
\end{equation}
where ${\rm Sc} = 1$ is the Schmidt number taken as 1 for simplicity \citep[see also][]{Johansen05}. Taking $w$ as the local gas disk scale height $H \equiv c_s / \Omega$, resolved dust rings can be leveraged to obtain the local ratio St/$\alpha$.

How can we break the degeneracy and separately determine St and $\alpha$? \citet{Rosotti20} computed gas rotation curves from $^{12}$CO emission for two ringed disks HD 163296 and AS 209 to obtain a precise value of St/$\alpha$ and placed an upper limit on $\alpha$ assuming the maximum St to be set by fragmentation (i.e., any larger grains would be ground down by relative motion between the grains set by their coupling to the gas turbulent motion). The fragmentation velocity however is highly uncertain and depends on not only the size of the grains but also their chemical composition and porosity \citep[e.g.,][]{Musiolik19,Kimura20}. Here, we demonstrate how, in addition to the dust ring width, dust ring mass can be used to solve for St and $\alpha$ separately.

The paper is organized as follows. Section \ref{sec:method} outlines our toy model of computing St and $\alpha$ and how we determine where or not a given dust ring is able to nucleate planetary mass objects. Section \ref{sec:results} presents our results, and their implications are summarized in Section \ref{sec:discussion}. We conclude in Section \ref{sec:concl}.

\section{Method} \label{sec:method}

We build a simple toy model of the dust mass growth in local pressure traps using the equation of motion of dust particle within a gaseous protoplanetary disk. Combining this equation with the measured properties of the observed rings, we break the degeneracy between $\alpha$ and ${\rm St}$.

\subsection{Disk profile} \label{ssec:disk-profile}

All the well-characterized dust rings are located at wide orbits beyond $\sim$10 AU where the disk heating is irradiation-dominated. Following \citet{Dullemond18}, we adopt a simple flared irradiated disk profile for the midplane temperature:
\begin{equation}
    T = \left(\frac{\phi L_\star}{8\pi a^2 \sigma_{\rm sb}}\right)^{1/4}
    \label{eq:Tdisk}
\end{equation}
where $L_\star$ is the measured luminosity of the central star, $\sigma_{\rm sb}$ is the Stefan-Boltzmann constant, and $\phi$ is the grazing angle of the incident starlight which range from $\sim$0.1 to $\sim$0.4 between 10 and 200 AU according to equations 14b and 14d of \citep{Chiang97}; see also equation 1 and Figure 3 of \citet{Dullemond01}---we take $\phi = 0.25$.

The shape of the disk surface density profile is less certain owing to uncertainty in the mechanism governing angular momentum loss/redistribution and the difficulty of obtaining well-resolved gas line measurements for many of the protoplanetary disks. We adopt $\Sigma_g \propto a^{-1}$ motivated by the gas emission line measurements of TW Hya by \citet{Zhang17} and the MAPS survey \citep[see][their Table 2]{Zhang21}. 

\subsection{Dust motion} \label{ssec:dust_motion}
The radial motion of a dust particle in the inertial frame can be written as
\begin{align} 
\label{eq:vdrift}
v_{\rm drift} &= -\frac{3\nu}{2a} \frac{1}{1+{\rm St}^2} - 2 v_{\rm hw} \frac{\rm St}{1+{\rm St}^2} \nonumber \\
&\sim -\frac{3}{2}\left(\frac{c_s}{v_k}\right)^2 v_k \left(\alpha + \frac{2}{3}|\gamma|{\rm St}\right)
\end{align}
where the negative sign indicates inward motion (towards the central star), $\nu \equiv \alpha c_s H$ is the kinematic viscosity of the gas, $\mu=2.3$ is the local disk gas mean molecular weight, $v_k = \Omega a$ is the Keplerian velocity, $v_{\rm hw} \equiv -(c_s^2/2 v_k)(\partial \log P/\partial \log a) \equiv -(c_s^2/2 v_k)\gamma$ is the azimuthal headwind velocity, and $P$ is the local disk gas pressure. With our adopted disk temperature and density profiles, $\gamma = -11/4$. We have taken the approximation of ${\rm St} \ll 1$ to write the second line, which we find to be valid a posteriori.

The first term in the bracket of equation \ref{eq:vdrift} corresponds to the grain's coupling to the radial ``advective'' motion of the gas (while to be more precisely, this is really a diffusive term for the gas but we assume $v_{\rm drift}$ to be always evaluated sufficiently inside the characteristic radius of the disk so that the gas motion under turbulent viscosity is inward rather than outward) and the second term in the bracket is the aerodynamic drag term. We have taken the drag term to also be inwardly-directed as is expected for a typical smooth disk. We emphasize that equation \ref{eq:vdrift} is intended to capture the motion of the dust grains through the unperturbed region of the disk before the grains are trapped.

\begin{figure}
    \centering
    \includegraphics[width=0.45\textwidth]{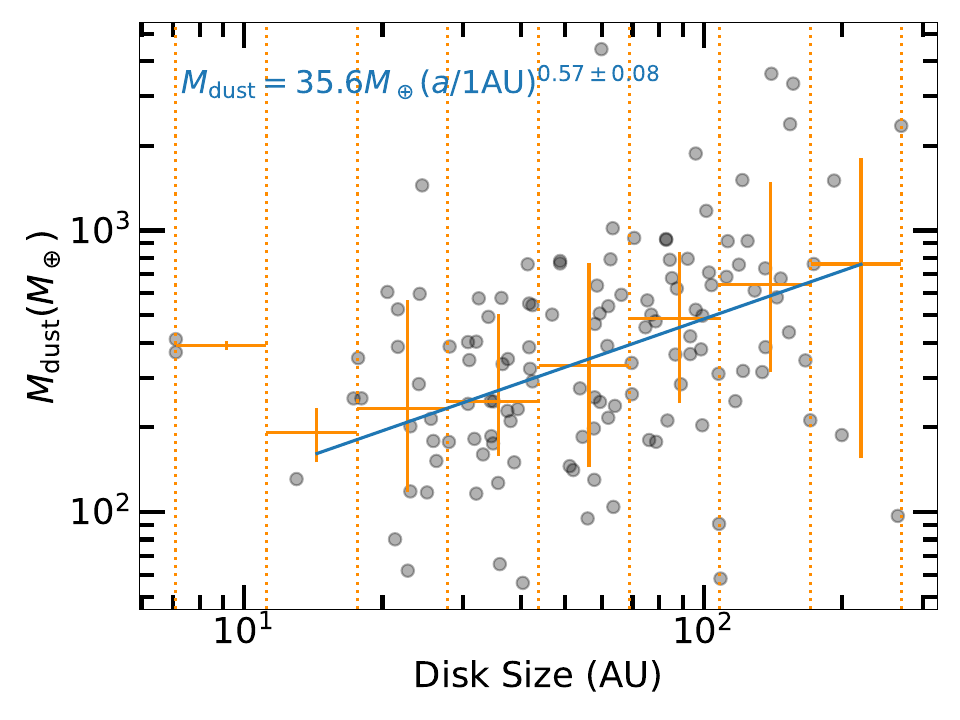}
    \caption{Total dust mass vs.~disk size for Class 0/I disks in the Orion star forming region with both ALMA (0.87 mm) and VLA (9 mm) measurements, denoted with grey circles. Vertical dotted lines delineate the boundaries of the 8 logarithmically spaced radial bins from the minimum to maximum disk sizes. Median $M_{\rm dust}$ for each bin is at the center of the orange crosshair with the lower and the upper tips of the crosshair indicating 16th and 84th percentiles, respectively. Least-squares fit accounting for the spread in $M_{\rm dust}$ in each bin (taken as the quadrature sum of the lower and upper spread) are shown in blue line with the result of the fit annotated at the top in blue text. Only the outer 6 radii bins are included in the fit as the inner two bins contain less than 3 points.}
    \label{fig:Mdust_disk}
\end{figure}

We then posit that there is some local perturbation in the gas pressure, without specifying the cause of it as it is irrelevant for our calculation. The dust grains from orbits outside of this perturbation will drift in following equation \ref{eq:vdrift} and be trapped into an axisymmetric ring. To find an expression for the amount of mass that is expected to be trapped within this ring, we first empirically derive the radial profile of initial dust mass budget using the mass and radius measurements of Class 0/I disks in the Orion  provided by \citet{Tobin20}, where both the ALMA (0.87 mm) and VLA (9 mm) measurements are available. Following the procedure of \citet{Chachan22}, we take the mass measurements from VLA (because disks are expected to be more optically thin at longer wavelengths) and the size measurements of ALMA---because disks are observed to be more compact at longer wavelengths which could be physical but also could be an effect of wavelength-dependent optical depths \citep[see, e.g.,][]{Tazzari16,Tripahi18}. Figure \ref{fig:Mdust_disk} illustrates the data and our best-fit initial dust mass profile using least-squares fit of the form $M_{\rm dust} = M_0 (a/{\rm 1\,AU})^\beta$ accounting for the spread in the data in each bin. Fitting through the median $M_{\rm dust}$ in each radial bin, we derive $M_{\rm dust} \propto a^{0.57\pm 0.08}$ which is steeper than what was reported by \citet{Lee22} but still agrees within 1-$\sigma$. While in Figure \ref{fig:Mdust_disk}, we display only the fit to the median $M_{\rm dust}$, in our analysis, we explore the full range from 5th to 100th percentiles of $M_{\rm dust}$ distribution, whereby $M_0$ and $\beta$ are empirically fit to the $M_{\rm dust}$ values of a given percentile in each radial bin.

We now write an expression for the total amount of dust that is expected to be trapped within a ring at a radius $a_{\rm ring}$ at time $t$ which is simply the amount of dust mass that exists between $a_{\rm ring}$ and some outer radius $a_0 > a_{\rm ring}$ from which the dust grains originate, corrected for some trapping efficiency:
\begin{equation}
    %M_{\rm ring}(t) = 36.3\,M_\oplus \epsilon_{\rm trap} \left[\left(\frac{a_0(t_{\rm age}-t)}{\rm 1 AU}\right)^{0.57}-\left(\frac{a_{\rm ring}}{\rm 1 AU}\right)^{0.57}\right]
    M_{\rm ring}(t) = M_0 \epsilon_{\rm trap} \left[\left(\frac{a_0(t_{\rm age}-t)}{\rm 1 AU}\right)^\beta-\left(\frac{a_{\rm ring}}{\rm 1 AU}\right)^\beta\right]
    \label{eq:Mring}
\end{equation}
where $t_{\rm age}$ is the system age, and $\epsilon_{\rm trap} \leq 1$ is the efficiency at which a given pressure perturbation can trap dust particles. This efficiency is found to be $\sim$0.6--0.8 by \citet{Lee22} through direct numerical simulations (see their Figure 4) but it is expected to vary with time, the properties of the pressure bump and the ratio between ${\rm St}$ and $\alpha$ \citep[see, e.g.,][their Section 4.4]{Zhu12}. Verifying with systematic investigations in 2D and 3D hydrodynamic simulations is an ongoing effort. For simplicity, we take a constant value $\epsilon_{\rm trap} = 0.6$. 

In equation \ref{eq:Mring}, we use $t_{\rm age}-t$ as the input time for $a_0$ to clarify that we are going back in time to locate the origin site of dust grains: i.e., $a_0$ is defined by integrating equation \ref{eq:vdrift} back in time $t$ with the boundary condition of $a_0(t_{\rm age}) = a_{\rm ring}$. From equation \ref{eq:Tdisk}, $c_s^2 \propto T \propto a^{-1/2}$ and because $v_k \propto a^{-1/2}$, $(c_s/v_k)^2 v_k$ in equation \ref{eq:vdrift} is spatially constant. If we further assume $\alpha$, $|\gamma|$, and ${\rm St}$ to be spatially constant:
\begin{equation}
    a_{0,{\rm St}}(t') = a_{\rm ring} + \frac{3}{2}\left(\frac{c_{s,r}}{v_{k,r}}\right)^2 v_{k,r} \left(\alpha + \frac{2}{3}|\gamma|{\rm St}\right)(t_{\rm age}-t'),
    \label{eq:a0_fixSt}
\end{equation}
where the subscript `r' clarifies $c_s$ and $v_k$ to be evaluated at the location of the observed ring and the subscript `St' specifies spatially constant St. Note that with our definitions, $M_{\rm ring}(t=0) = 0$ (no grains have drifted in yet) and $M_{\rm ring}(t_{\rm age})$ is maximized (maximum time elapsed).

The initial drift of dust grains into the pressure bump is considered to occur in a smooth disk so we maintain the assumption of spatially constant $\alpha$ and $|\gamma|$. Assuming a constant ${\rm St}$ is less certain as this number depends on the disk gas surface density and temperature profile even for a fixed grain size. We now derive $a_0(t)$ assuming a fixed grain size. Again, since we are in the outer disk, the gas mean free path is relatively large and the mm-cm sized grains are most likely in Epstein drag regime,\footnote{Using the minimum mass extrasolar nebula \citep{Chiang13} and the gas mean free path $\lambda = 4\times 10^{-9}/\rho_g$ where $\rho_g$ is the volumetric mass density of disk gas, $\lambda \gtrsim 200$ cm beyond 10 AU.} so from equation \ref{eq:St}, 
\begin{equation}
    {\rm St} = \frac{\rho_s s}{\Sigma_g}
    \label{eq:St_rewrite}
\end{equation}
where $\Sigma_g = \rho_g H$ is the surface density of the disk gas. Given that $\Sigma_g \propto a^{-1}$, ${\rm St} \propto a$ for a fixed $s$. Integrating equation (\ref{eq:vdrift}) back in time with ${\rm St} \propto a$ with the same boundary condition $a_0(t_{\rm age}) = a_{\rm ring}$:
\begin{align}
    \frac{a_{0,s}(t')}{a_{\rm ring}} = &-\frac{3}{2}\frac{\alpha}{|\gamma|{\rm St_r}} \nonumber \\
    &+\left(1 + \frac{3}{2}\frac{\alpha}{|\gamma|{\rm St_r}}\right){\rm Exp}\left[\frac{c_{s,r}^2}{v_{k,r}}|\gamma|{\rm St_r}\frac{(t_{\rm age}-t')}{a_{\rm ring}}\right],
    \label{eq:a0_fixs}
\end{align}
where the subscript `s' denotes spatially constant grain size.

When a system features two rings, we account for potential reduction of the dust masses that can reach the inner ring owing to the partial filtering by the outer ring as follows. First, we check if $a_0$ of the inner ring is within $a_{\rm ring}$ of the outer ring, and if so, the initial dust sourcing the inner ring is unaffected by the outer ring. If the condition is not met however, we account for the filtering by the outer ring by re-computing the effective $M_{\rm 0,new}$ for the inner ring:
\begin{equation}
    M_{\rm 0,orig}\left(\frac{a_{\rm 0,out}}{1\,{\rm AU}}\right)^\beta - M_{\rm ring,out} = M_{\rm 0,new}\left(\frac{a_{\rm 0,out}}{1\,{\rm AU}}\right)^\beta
\end{equation}
where $M_{\rm 0,orig}$ is the original unaffected $M_0$, $a_{\rm 0,out}$ is the origin site of dust grains that source the {\it outer} ring, and $M_{\rm ring,out}$ is the measured dust mass of the outer ring. We then recompute St and $\alpha$ of the inner ring using the new $M_{\rm 0,new}$. 

\begin{deluxetable*}{ccccccccccc}
\tablecaption{Well-resolved dust rings in the DSHARP survey drawn from \citet{Dullemond18}.}
\tablehead{\colhead{Source} & \colhead{Name} & \colhead{Age} & \colhead{$M_\star$} & \colhead{$L_\star$} & \colhead{$a_{\rm ring}$} & \colhead{$w_d$} & \colhead{$M_{\rm ring}$} & \colhead{(St$/\alpha)_{\rm min}$} & \colhead{(St$/\alpha)_{\rm max}$} & \colhead{(St$/\alpha)_{\rm rot}$} \\ 
\colhead{} & \colhead{} & \colhead{(log (yr))} & \colhead{($M_\odot$)} & \colhead{($L_\odot$)} & \colhead{(au)} & \colhead{(au)} & \colhead{($M_\oplus$)} & \colhead{} & \colhead{} & \colhead{} } 
\startdata
AS 209 & B74 & 6.0 & 0.83 & 1.41 & 74.2 & 3.38 & 31.5 & 4.54 & 7.14 & 5.56 \\
AS 209 & B120 & 6.0 & 0.83 & 1.41 & 120.4 & 4.11 & 69.8 & 6.67 & 9.09 & 7.69 \\
Elias 24 & B77 & 5.3 & 0.78 & 6 & 76.7 & 4.57 & 40.8 & 3.75 & 12.99 & 0 \\
HD 163296 & B67 & 6.9 & 2.04 & 17 & 67.7 & 6.84 & 56.0 & 3.85 & 5.00 & 4.35 \\
HD 163296 & B100 & 6.9 & 2.04 & 17 & 100.0 & 4.67 & 43.6 & 20.0 & 33.3 & 25.0 \\
GW Lup & B85 & 6.3 & 0.46 & 0.33 & 85.6 & 4.8 & 37.0 & 0 & 3.23 & 0 \\
HD 143006 & B41 & 6.6 & 1.78 & 3.8 & 41.0 & 3.9 & 9.9 & 0 & 5.56 & 0 \\
HD 143006 & B65 & 6.6 & 1.78 & 3.8 & 65.2 & 7.31 & 25.6 & 0 & 0.91 & 0 \\
\enddata
\tablecomments{Column 1: name of the source. Column 2: name of the ring. Column 3: system age taken from Table 1 of \citet{Andrews18}, with the exception of HD 163296 for which we take the value quoted in the original reference \citep{Fairlamb15}. Column 4: mass of the host star. Column 5: luminosity of the host star. Column 6: measured orbital distance of the ring. Column 7: underlying width of the ring after deconvolution. Column 8: measured dust mass of the ring including the correction for the optical depth ($M^{\rm true}_{\rm d}$ of \citealt{Dullemond18}). Column 9: minimum inferred value of St$/\alpha$ (inverse of column 12 of \citealt{Dullemond18}, their Table 3). Column 10: maximum inferred value of St$/\alpha$ (inverse of column 11 of \citealt{Dullemond18}, their Table 3). Column 11: measured St$/\alpha$ with gas rotation curves from \citet{Rosotti20}. For rings re-analyzed by \citet{Rosotti20}, we replace the values of columns 9 and 10 with the lower and upper 1-$\sigma$ error limits based on the measurements of \citet{Rosotti20}. The corresponding ratio ${\rm St}/\alpha$ is recorded as zero when it is not available. For rings without gas rotation curve measurements, we update (St/$\alpha$)$_{\rm min}$ with our $\phi = 0.25$ as compared to $\phi = 0.02$ adopted by \citet{Dullemond18} which increases the gas disk scale height by $\sim$37\% and boosts the derived St/$\alpha$ by order unity factors. For GW Lup B85, this correction drives minimum St/$\alpha$ to be slightly larger than the maximum St/$\alpha$ and so we adopt the maximum value to be the true value and set the minimum value to zero. For HD 143006 rings, the new gas disk scale heights remain smaller than $w_d$ and so (St/$\alpha$)$_{\rm min}$ are recorded as zero.}
\vspace{-1.2cm}
\label{tab:dsharp-disk}
\end{deluxetable*}

\begin{deluxetable*}{cccccccc}
\tablecaption{Limits on St and $\alpha$ for dust rings in Table \ref{tab:dsharp-disk}}
\tablehead{\colhead{Source} & \colhead{Name} & \colhead{$f(>M_{\rm dust})_Q$} & \colhead{${\rm St}_Q$} & \colhead{$\alpha_Q$} & \colhead{$f(>M_{\rm dust})_M$} & \colhead{${\rm St}_M$} & \colhead{$\alpha_M$} \\ 
\colhead{} & \colhead{} & \colhead{(percentile)} & \colhead{} & \colhead{} & \colhead{(percentile)} & \colhead{} & \colhead{}} 
\startdata
AS 209 & B74 & 72.54 & 4.50E-04 & 8.10E-05 & 30.91 & 2.03E-03 & 3.66E-04 \\
AS 209 & B120 & 72.54 & 1.22E-03 & 1.59E-04 & 30.91 & 5.01E-03 & 6.51E-04 \\
Elias 24 & B77 & 100 & 9.21E-04 & 2.46E-04 & 15.56 & 1.60E-02 & 4.26E-03 \\
HD 163296 & B67 & 29.48 & 3.54E-04 & 8.14E-05 & 27.07 & 4.73E-04 & 1.09E-04 \\
HD 163296 & B100 & 29.48 & 3.54E-04 & 1.42E-05 & 27.07 & 4.56E-04 & 1.82E-05 \\
GW Lup & B85 & 34.45 & 9.31E-04 & 2.88E-04 & 5 & 2.62E-03 & 8.12E-04 \\
HD 143006 & B41 & 34.05 & 9.62E-05 & 1.73E-05 & 5 & 3.08E-04 & 5.55E-05 \\
HD 143006 & B65 & 34.05 & 2.30E-04 & 2.53E-04 & 5 & 6.72E-04 & 7.38E-04 \\
\enddata
\tablecomments{Column 1: name of the source. Column 2: name of the ring. Column 3: maximum $M_{\rm dust}$ percentile for gravitational stability ($Q_T \geq 1$). Column 4: minimum St for $Q_T \geq 1$. Column 5: corresponding $\alpha$ for St$_Q$. Column 6: minimum $M_{\rm dust}$ percentile to explain ring dust mass. Column 7: maximum St. Column 8: corresponding $\alpha$ for maximum St.}
\label{tab:ring-analy}
\end{deluxetable*}

\subsection{Solving for St and $\alpha$} \label{ssec:solve-St-alpha}

Substituting equations (\ref{eq:a0_fixSt}) or (\ref{eq:a0_fixs}) into equation (\ref{eq:Mring}) and equating it to the measured total dust mass in an observed ring $M_{\rm ring}$ at the measured ring location $a_{\rm ring}$ and at the estimated age of the host system $t_{\rm age}$, we can solve for a quantity that is a combination of $\alpha$ and ${\rm St}$ which can be separated by using the measured ${\rm St}/\alpha$ from, e.g., \citet{Dullemond18}; see Table \ref{tab:dsharp-disk} for the list of DSHARP disks with well-resolved rings on which we focus our analysis. In case of spatially constant ${\rm St}$,
\begin{equation}
    \alpha + \frac{2}{3}|\gamma|{\rm St} = \frac{\left(\frac{M_{\rm ring}}{M_0\epsilon_{\rm trap}}+\left(\frac{a_{\rm ring}}{{\rm 1 AU}}\right)^\beta\right)^{1/\beta}-\frac{a_{\rm out}}{\rm 1 AU}}{\frac{3}{2}\left(\frac{c_{s,r}}{v_{k,r}}\right)^2v_{k,r}t_{\rm age}/{\rm 1 AU}}.
\end{equation}
All the quantities on the right hand side are either measured or specified in our model setup and so is $\gamma$ so we can calculate the left hand side and combine it with the measured ${\rm St}/\alpha$ to separately determine $\alpha$ and St. 

In case of spatially constant $s$, we can directly solve for ${\rm St_{\rm r}}$:
\begin{align}
    {\rm St}_{\rm r} = &\left(\frac{v_{k,r}}{c_{s,r}^2}\right)\frac{a_{\rm ring}}{t_{\rm age}|\gamma|} \nonumber \\
    &\times\log\left(\frac{\left(\frac{M_{\rm ring}}{M_0\epsilon_{\rm trap}}+\left(\frac{a_{\rm ring}}{{\rm 1 AU}}\right)^\beta\right)^{1/\beta} + \frac{3}{2}\frac{\alpha}{|\gamma|{\rm St}_r}a_{\rm ring}}{\left(\frac{a_{\rm ring}}{\rm 1 AU}\right)+\frac{3}{2}\frac{\alpha}{|\gamma|{\rm St_r}}\left(\frac{a_{\rm ring}}{\rm 1 AU}\right)}\right) \nonumber \\
\end{align}
where all values on the right hand side are known or measured (and the $\log$ here is natural logarithm). We can then compute $\alpha$ using the measured ${\rm St}/\alpha$.

Above analysis implicitly assumes the $\alpha$ that parametrizes the diffusion of dust mass in the pressure bump (equation \ref{eq:dust-ring-width}) is equivalent to that parametrizes the angular momentum transfer of the disk gas (equation \ref{eq:vdrift}). Our approach here is valid as long as the effective values of the two parametrization are similar, confirmation of which would depend on the exact source of both the pressure bump and the redistribution of disk gas angular momentum. We defer such confirmation to future numerical studies.

\begin{figure*}
    \centering
    \includegraphics[width=\textwidth]{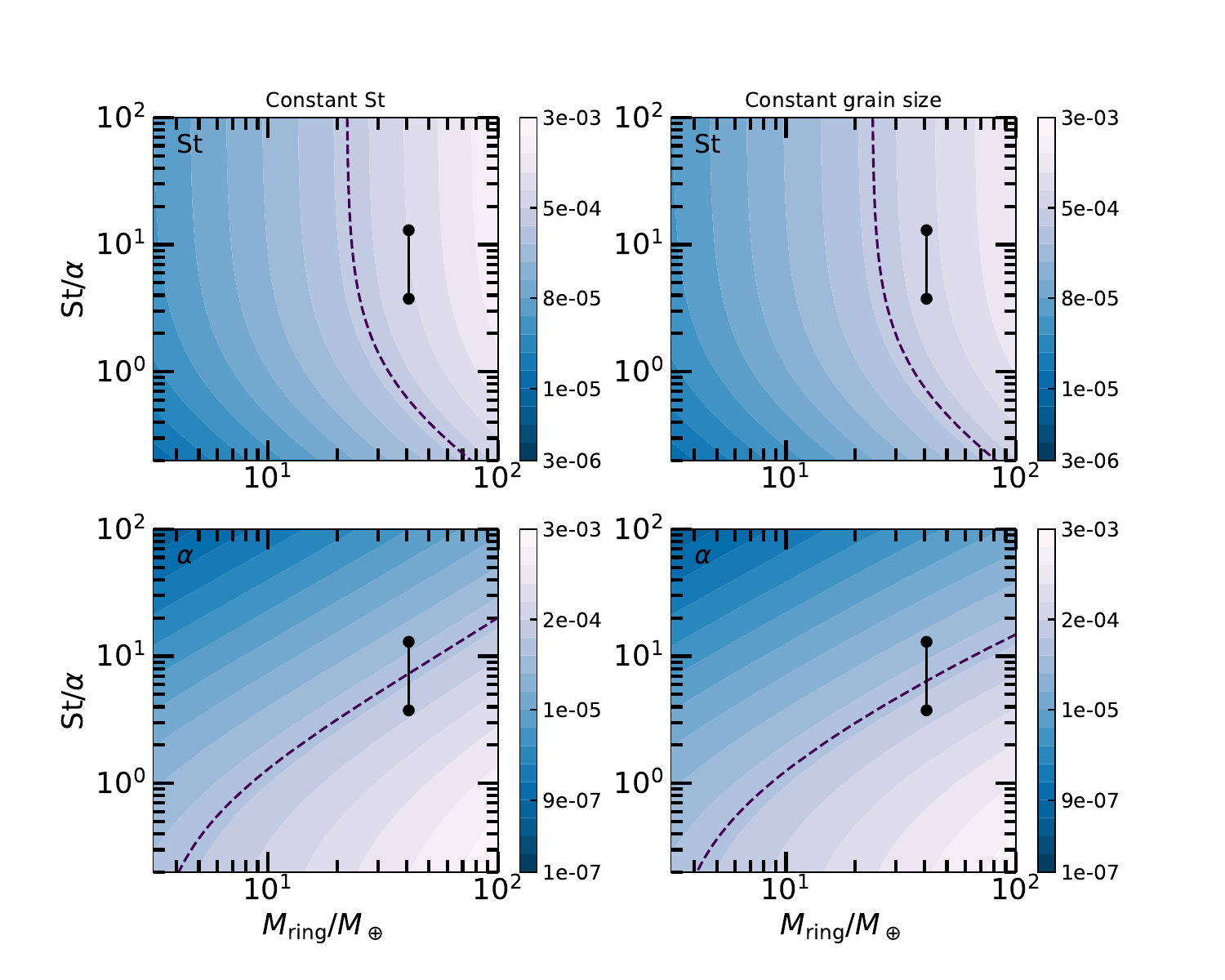}
    \caption{Contour map of St (top row) and $\alpha$ (bottom row) as a function of ${\rm St}/\alpha$ and the ring mass for a system parameter following Elias 24 B77 (see Table \ref{tab:dsharp-disk}) with median $M_{\rm dust}$ (see Figure \ref{fig:Mdust_disk}). Whether we assume constant St (left column) or constant grain size (right column) makes negligible difference in the calculated St and $\alpha$. The purple dashed lines in the top row draw the St that corresponds to Toomre $Q = 1$ such that larger St would be consistent with local gravitational stability (see Section \ref{ssec:limits}). The purple dashed line in the bottom row delineates $\alpha = 10^{-4}$ below which the disk is in the strongly inviscid limit. The measured St/$\alpha$ and $M_{\rm ring}$ for Elias 24 B77 are indicated with black circles with the upper and lower limits connected by a line.}
    \label{fig:StA_Mring_map}
\end{figure*}

We illustrate in Figure \ref{fig:StA_Mring_map} the overall trend in the inferred St and $\alpha$ for a range of ${\rm St}/\alpha$ and $M_{\rm ring}$, showing Elias 24 B77 as an example case. The Stokes number is largely insensitive to the ratio ${\rm St}/\alpha$ and more of a function of $M_{\rm ring}$ except when $\alpha \gtrsim {\rm St}$. Larger ring mass translates into larger St because larger amount of mass needs to drift into a given ring location for a fixed total dust budget (equivalent behavior is observed when $M_{\rm dust}$ is lower for a fixed ring mass). The St-$M_{\rm ring}$ relation flattens out at lower ${\rm St}/\alpha$ because the dust radial motion becomes more dominated by their coupling to the gas advection.

Compared to the inferred St, the turbulent $\alpha$ varies with both ${\rm St}/\alpha$ and $M_{\rm ring}$, increasing overall with lower ${\rm St}/\alpha$ and higher $M_{\rm ring}$, with the $\alpha$-$M_{\rm ring}$ relation strengthening (and $\alpha-{\rm St}/\alpha$ relation weakening) at lower ${\rm St}/\alpha$. The overall increase in $\alpha$ with more massive ring can be understood as due to the need to drift more dust into a ring location, identical to the cause for the positive St-$M_{\rm ring}$ relation. The rise in $\alpha$ for lower St/$\alpha$ down to St/$\alpha \sim 1$ appears because for a fixed St, lower St/$\alpha$ implies higher $\alpha$. Below St/$\alpha \lesssim 1$, the $\alpha-M_{\rm ring}$ relation strengthens because dust coupling to the gas advection dominates dust drift (the same effect reducing St-$M_{\rm ring}$ relation).

Comparing the two columns of Figure \ref{fig:StA_Mring_map}, the aforementioned observed behaviors as well as the magnitude of St and $\alpha$ appear generally identical between constant St and constant grain size models, demonstrating the robustness of our calculation to the assumptions of microphysics of dust grains. Even though the physics of dust-gas dynamics is controlled by St, the images from ALMA observations are sensitive to grain size and so for the remainder of the paper, we will focus our attention on our constant grain size models.

\subsection{Placing limits on St and $\alpha$} \label{ssec:limits}

Following the calculations outlined in Sections \ref{ssec:dust_motion} and \ref{ssec:solve-St-alpha}, we find that the derived St and $\alpha$ vary most sensitively to $M_0$ and $\beta$, which are controlled by the chosen percentile of $M_{\rm dust}$ distribution (see Figure \ref{fig:Mdust_disk}). Here, we describe how we constrain the lower and upper limits on reasonable $M_{\rm dust}$ percentiles, which in turn correspond to upper and lower limits, respectively, on the derived St and $\alpha$.

First, we choose which St/$\alpha$ to adopt. When the rotation curve measurements are available (AS 209 and HD 163296 rings), the width of the gas pressure bump can be more precisely determined so we always use (St/$\alpha$)$_{\rm rot}$. In the absence of such measurements, we have a choice between the minimum and maximum St/$\alpha$. The minimum value is evaluated using equation \ref{eq:dust-ring-width} by taking $w=H$ whereas the maximum value is derived by setting $w$ as the separation between rings in a given system or the distance between the dust ring to the nearest minimum (see Table 3 caption of \citealt{Dullemond18} for more detail). We favor the minimum value where available because it is more self-consistent with the setup of our model.

Next, we establish that with lower $M_{\rm dust}$ and at fixed St/$\alpha$, the derived St is higher, as seen in Figure \ref{fig:St_Md0_ring} in the Appendix, and it follows that $\alpha$ is also higher. Such a trend arises because in disks lighter with dust, larger radial extent ($a > a_{\rm ring}$) of dust must be sourced to account for a given $M_{\rm ring}$ and for given system age, necessitating more rapid inward drift of dust.

We now proceed to establish limits on $M_{\rm dust}$ percentiles. The lower limit is given by when the corresponding origin site of the dust $a_0$ for either the inner or the outer ring exceeds 300 AU, which we consider to be unphysically large---this is the maximum size among the Class 0/I disks shown in Figure \ref{fig:Mdust_disk}. Any smaller $M_{\rm dust}$ is considered incompatibly low with the measured disk conditions. In cases where $M_{\rm dust}$ remains consistent with a given ring dust mass down to 5th percentiles, we take the 5th percentile as the lower limit. For all such cases, St appears flat for $\lesssim$10th percentile so we set this 5th percentile as the absolute lower limit (see Figure \ref{fig:St_Md0_ring}).

The upper limit on $M_{\rm dust}$ percentile is given by the requirement for the underlying gas disk to be gravitationally stable, which should locally satisfy
\begin{equation}
    Q_T \equiv \frac{c_s \Omega}{\pi G\Sigma_g} = \frac{c_s \Omega}{\pi G\rho_s s}{\rm St} > 1
\end{equation}
where the second equality derives from equation \ref{eq:St_rewrite}. For our analysis, we adopt $\rho_s = 1\,{\rm g\,cm^{-3}}$ and $s=\lambda/2\pi$ where $\lambda = 1.25$mm is the wavelength of ALMA observations \citep{Andrews18}. It follows that the local St for a given ring has a lower limit:
\begin{equation}
    {\rm St} > \frac{\pi G \rho_s s}{c_s\Omega}
\end{equation}
while keeping in mind the uncertainties in $\rho_s$ and $s$ likely result in at least order unity variations. In one case (Elias 24 B77; see Figure \ref{fig:St_Md0_ring}), we find that even at 100th percentile of $M_{\rm dust}$, $Q_T > 1$ and so we take 100 as the upper limit percentile.

For systems with more than one ring, we ensure the same $M_{\rm dust}$ percentile is used within a given system. Among the rings in a system, we take the smallest $M_{\rm dust}$ percentile for local gravitational stability (i.e., the global upper limit on $M_{\rm dust}$) and the largest $M_{\rm dust}$ percentile for dust mass budget (i.e., the global lower limit on $M_{\rm dust}$). Table \ref{tab:ring-analy} summarizes the derived limits on St and $\alpha$.

\begin{figure}
    \centering
    \includegraphics[width=0.5\textwidth]{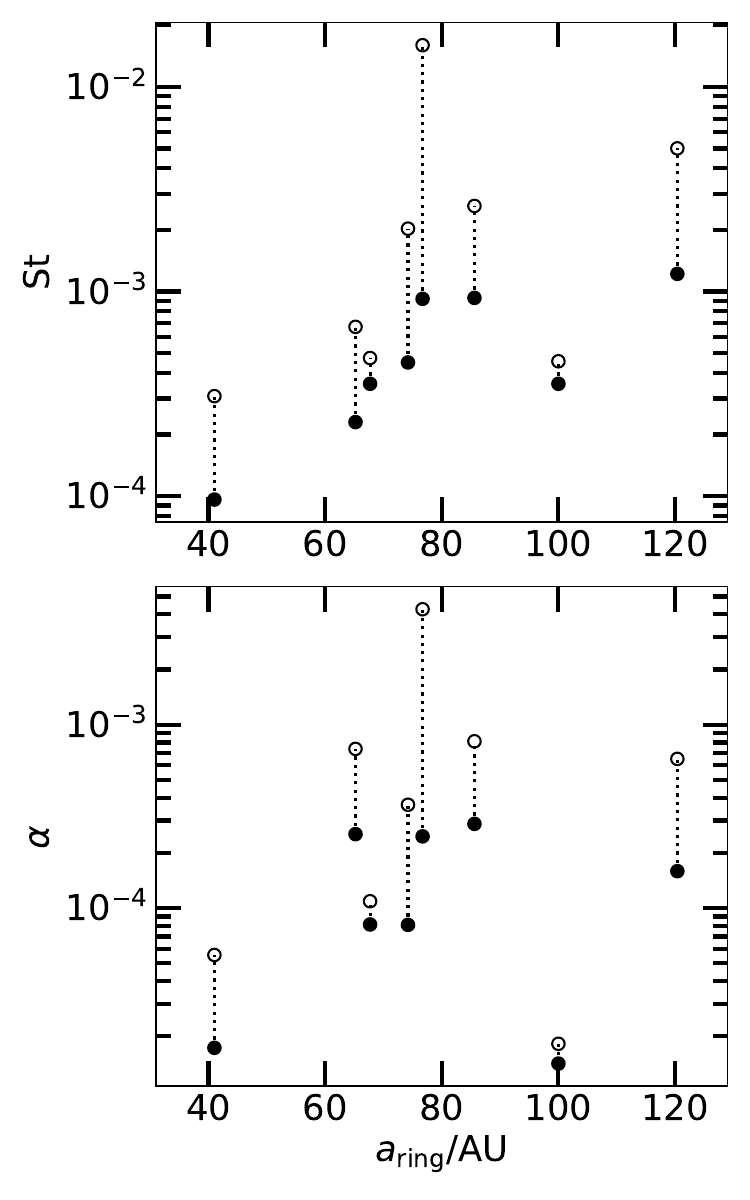}
    \caption{Derived St (top row) and $\alpha$ (bottom row) for individual rings from Table \ref{tab:dsharp-disk}. Filled circles represent minimum St for local gravitational stability and corresponding $\alpha$ whereas the unfilled circles represent maximum St that satisfies the disk mass budget and its corresponding $\alpha$.
    In general, all the rings indicate low St $\lesssim 10^{-2}$ and low $\alpha \lesssim 10^{-3}$, consistent with being approximately inviscid.}
    \label{fig:St_Alpha_sys}
\end{figure}

\begin{figure}
    \centering
    \includegraphics[width=0.5\textwidth]{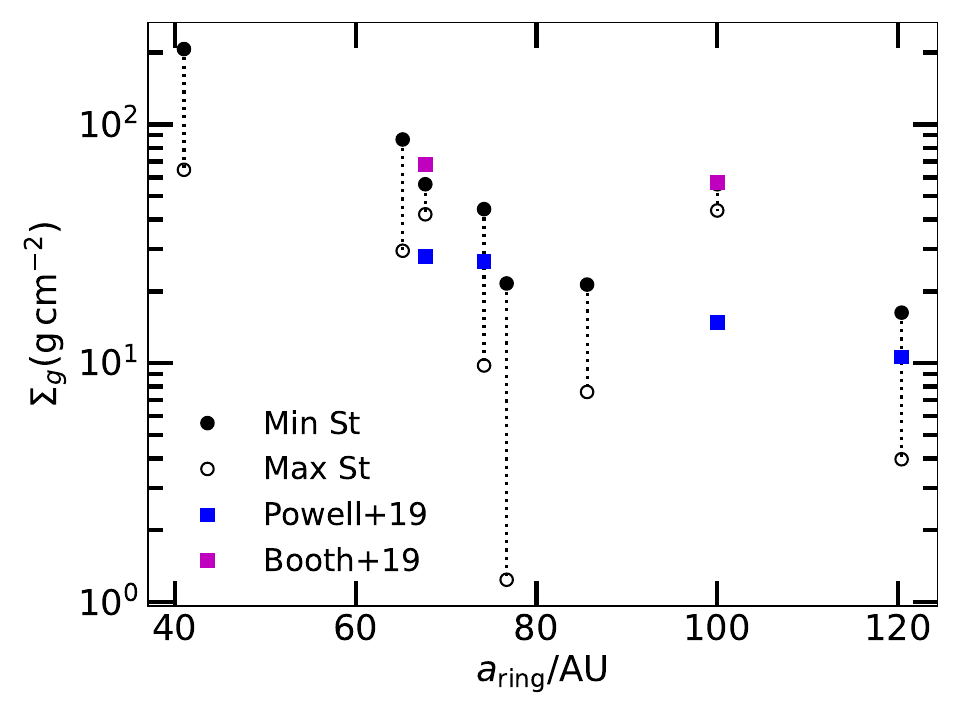}
    \caption{Estimated local gas surface density $\Sigma_g$ for derived minimum St for gravitational stability (St$_Q$ in Table \ref{tab:ring-analy}; filled circle) and maximum St to be consistent with ring dust mass (St$_M$ in Table \ref{tab:ring-analy}; open circle) for each ring. Where available, we indicate with blue squares the estimated $\Sigma_g$ from \citet[][AS 209 B74 and B120; HD 163296 B67 and B100]{Powell19} and with magenta squares the measured $\Sigma_g$ from \citet[][HD 163296 B67 and B100]{Booth19} where for the latter, we take the values in Table 2 of \citet{Rosotti20}.}
    \label{fig:Sigma_gas_St}
\end{figure}

\section{Results} \label{sec:results}

As illustrated in Figure \ref{fig:St_Alpha_sys}, all the rings studied in this paper feature low St $\lesssim 10^{-2}$ and low $\alpha \lesssim 10^{-3}$. The derived range in both St and $\alpha$ are within order unity with the exception of Elias 24 B77 because we can find solutions for this particular ring over the full range of $M_{\rm dust}$ (see Figure \ref{fig:St_Md0_ring}). As we will discuss more in Section \ref{sec:discussion}, such low St would slow down much of the effects of dust-gas dynamics including clumping and core coagulation. The low values of $\alpha$ we obtain are consistent with molecular line measurements of \citet{Flaherty17} and suggest the disks to be nearly inviscid, at least at the large orbital distances ($\gtrsim$10 AU) where we observe dust rings.

In Figure \ref{fig:Sigma_gas_St}, we compare the estimated range of $\Sigma_g$ based on our derived St to what is reported in the literature. For AS 209 and HD 163296 rings, \citet{Powell19} derive local $\Sigma_g$ assuming the local particle drift timescale to be equal to the system age \citep[see also][]{Powell17}. We find that their $\Sigma_g$ falls well within the range expected from our derivation for AS 209 B74 and B120. For HD 163296 rings, their $\Sigma_g$ are lower than even our lower limit. For this disk, a more direct measurement of gas disk exists with $^{13}$C$^{17}$O (J=3-2) line that is expected to be optically thin down to $\sim$50 AU \citep[see][their Figure 2d]{Booth19}. Using the disk model of \citet{Booth19}, \citet{Rosotti20} quote $\Sigma_g = 68\,{\rm g\,cm^{-2}}$ and $57\,{\rm g\,cm^{-2}}$ for B67 and B100 which we show in Figure \ref{fig:Sigma_gas_St}.\footnote{Reading Figure 3e Model 2 of \citet{Booth19} and converting the line intensity to gas surface density following the procedure of \citet{Carney19} with spectral information from \citet{CDMS} and $^{13}$C$^{17}$O to H$_2$ abundance ratio $5.39\times 10^{-10}$, we can recover $\Sigma_g$ that are close but higher than what are quoted in \citet{Rosotti20} by $\sim$20--40\% likely due to estimating the measured $^{13}$C$^{17}$O line intensity by eye.} These more directly measured $\Sigma_g$ are at the maximum value allowed for gravitational stability. In case of B67, the measured $\Sigma_g$ is even slightly higher; the difference however is less than a factor of 2 which can be easily absorbed into the uncertainties in the grain bulk density $\rho_s$, grain size $s$, and also the age of the system.

We close this section with a comment on the difference between \citet{Powell17} (and by extension \citealt{Powell19}) and our model. For the specific case of HD 163296, while the adopted stellar parameters are different, we do not think this is the main cause of the different $\Sigma_g$. \citet{Powell19} use 36$L_\odot$ and 2.3$M_\odot$ as compared to our 17$L_\odot$ and 2.04$M_\odot$. Combined with their choice of temperature profile (see their equations 3 and 4), their local headwind velocity is smaller than ours by $\sim$20\%. They also adopt younger age estimate for the system 5 Myr which is $\sim$37\% shorter than the age we use. These differences, however, are washed out by their adopting $\rho_s = 2\,{\rm g\,cm^{-3}}$ instead of our $\rho_s = 1\,{\rm g\,cm^{-3}}$ (c.f.~equation 1 of \citealt{Powell19}).

Functionally, the calculation of \citet{Powell17} is equivalent to removing the coupling to gas advection term in $v_{\rm drift}$ (equation \ref{eq:vdrift}), integrating it to some prescribed outer edge of the disk $r_{\rm edge}$ for a chosen grain size over the system age to solve for St:
\begin{equation}
    {\rm St} \sim \frac{r_{\rm edge}}{t_{\rm age}\frac{c_s^2}{v_k}|\gamma|}
\end{equation}
and then plugging this St into equation \ref{eq:St_rewrite} to solve for $\Sigma_g$. In this sense, our approach is more complete as we account for the dust coupling to inward gas flow which in turn requires us to solve for St and $\alpha$ that are also consistent with the measured dust mass within the dust ring in addition to the location of the dust ring. We conclude that the difference between our result and that of \citet{Powell19} mostly arises from our more complete treatment of dust motion and our introduction of extra constraint from dust mass within rings. We are further encouraged by the greater agreement between our derived $\Sigma_g$ and the direct gas measurements in HD 163296.

\section{Discussion} \label{sec:discussion}

While the overconcentration of dust in the observed rings potentially provides ideal sites of planet formation, the fact that we see these rings over the system age imply that they must be stable against vigorous formation of stable clumps and planetary mass objects. In this section, we discuss the implication of our derived St and $\alpha$ on the survival of rings and planet formation at wide orbital separations.

\subsection{Rings are Stable to Clumping by Streaming Instability} \label{ssec:feedback}

\begin{figure}
    \centering
    \includegraphics[width=0.5\textwidth]{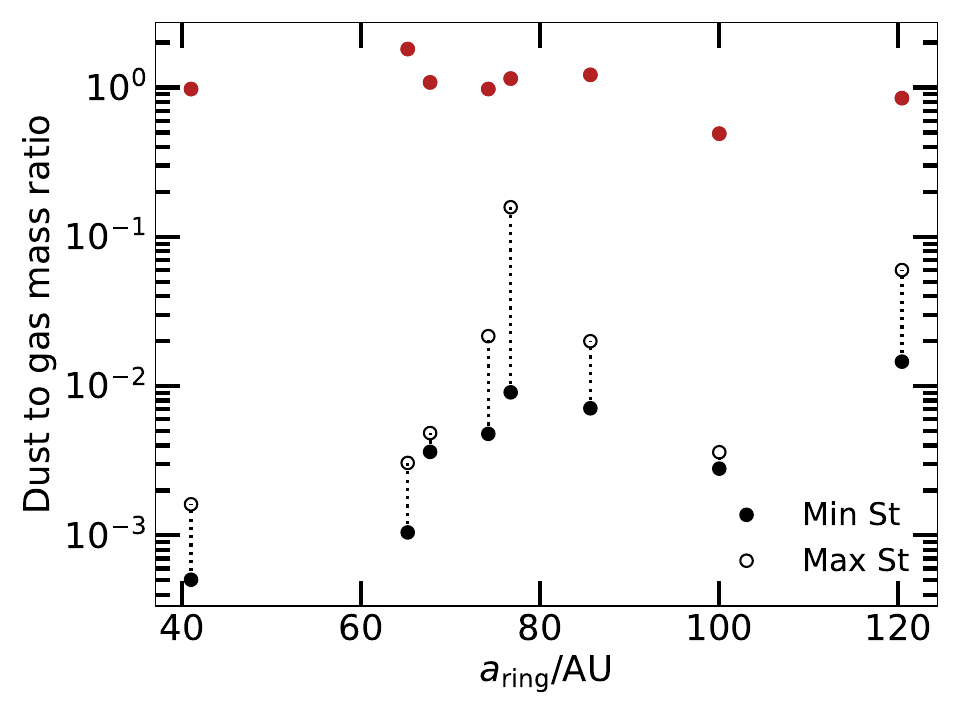}
    \caption{Estimated local dust-to-gas mass ratio for minimum St (filled circles) and maximum St (unfilled circles) for each ring. Plotted in red circles are the critical dust-to-gas mass ratio for triggering streaming instability taking into account extra gas turbulence (e.g., unknown source generating the dust trap) in addition to what would be generated from dust-gas interaction. All rings studied are expected to be stable against streaming instability.}
    \label{fig:d2g-ring}
\end{figure}

One oft-invoked mechanism to produce dust clumps is streaming instability \citep[SI; e.g.,][]{Youdin05} which arises from the non-zero drift of solids with respect to gas and the dust feedback on the motion of the gas. Key criteria for clump formation by SI is to have the local dust-to-gas mass ratio $\Sigma_{\rm ring}/\Sigma_g$ and St high enough.\footnote{Clumping is inherently nonlinear and so while (linear) SI always exists, we are interested in the conditions that allow for SI growth timescale that is shorter than the system age and more critically that are explicitly found to trigger clump formation.} We first compute the dust-to-gas ratio for the rings studied in this paper by converting $M_{\rm ring}$ of each ring into $\Sigma_{\rm ring}$ following (c.f. equation 8 of \citealt{Dullemond18})
\begin{equation}
    \Sigma_{\rm ring} \equiv \frac{M_{\rm ring}}{(2\pi)^{3/2}a_{\rm ring}w_d},
    \label{eq:Sig_ring}
\end{equation}
and $\Sigma_g$ follows from Figure \ref{fig:Sigma_gas_St}.

\citet{Li21} provide an empirical fit to their numerical calculations of the critical dust-to-gas mass ratio for clumping by SI. We adopt their equation 14 which takes into account extra turbulence in addition to that by dust-gas interaction that drives the instability to be consistent with our definition of dust ring (equation \ref{eq:dust-ring-width}). We emphasize that for our derived St and $\alpha$ the additional turbulence term that sets the dust disk scale height ($H_{\rm p,\alpha}$ in \citealt{Li21}) dominates over that expected from pure dust-gas interaction (their $H_{\rm p,\eta}$). Figure \ref{fig:d2g-ring} demonstrates that for all the rings we study, the derived dust-to-gas mass ratio is always smaller than that required for clumping by SI.

It is not surprising that accounting for the extra turbulence impedes SI. Dust feedback onto gas is most effective when the relative local concentration of dust is already comparable to that of gas. Higher level of turbulence puffs up the solid disk diluting the dust concentration in the midplane, which in turn would lessen the effect of dust feedback. In fact, \citet{Chen20} find that when $\alpha \gtrsim {\rm St}^{1.5}$, the growth timescale by SI becomes untenably long (defined as $\sim$10$^6$ times local orbital time; see also \citealt{Umurhan20}) and we verify that this inequality is met for all the rings we study with the exception of Elias 24 B77 at its maximum St and corresponding $\alpha$ (but even for this case, $\alpha/{\rm St}^{1.5} \sim 0.6$).

Faced with the difficulties with clump formation by SI in turbulent rings, \citet{Liu23} identify dust-driven Rossby wave instability that may lead to clumping. Although intriguing, we cannot verify whether these new instabilities would be active in the rings we study because \citet{Liu23} do not report results for sufficiently low dust-to-gas mass ratio. We therefore conclude that the rings studied here are stable against clumping by SI but proceed with a case where initial clumps may nevertheless emerge by other dust-driven instabilities.

\subsection{Dust Clumps in Rings are Unstable} \label{ssec:core-coag}

\begin{figure}
    \centering
    \includegraphics[width=0.5\textwidth]{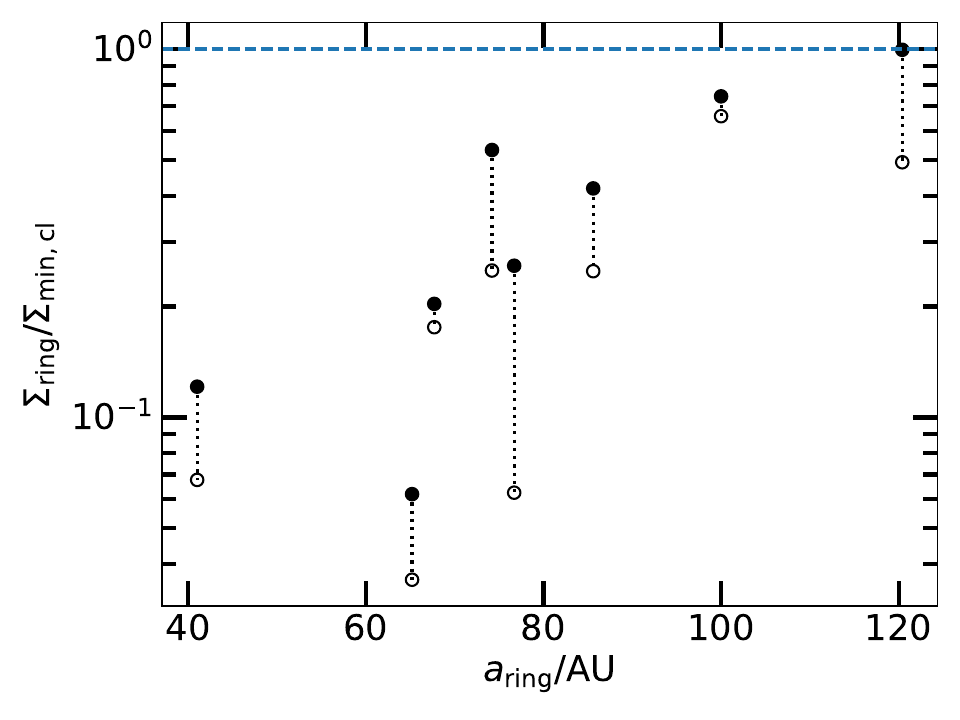}
    \caption{The ratio of ring surface density (equation \ref{eq:Sig_ring}) to the minimum density required for nucleation of bound clump (equation \ref{eq:Sig_ring_plt}) vs.~the radial location for each ring. Plotting scheme is identical to that of Figure \ref{fig:St_Alpha_sys}. None of the rings are able to nucleate stable clumps.}
    \label{fig:Sigma_diag}
\end{figure}

If rings were to create planetary cores, the initial dust clumps must be gravitationally bound against particle diffusion \citep{Klahr18} and stable against tidal shear \citep{Gerbig20}. Clump formation that meets both requirements is possible when 
\begin{equation}
    \Sigma_{\rm ring} > \left(\frac{M_\star v_{\rm rms}^2}{\pi^2 G a_{\rm ring}^3}\right)^{1/2} = \left(\frac{M_\star}{\pi^2 G a_{\rm ring}^3}\right)^{1/2} c_s \left(\frac{\alpha}{1+{\rm St}}\right)^{1/2}
    \label{eq:Sig_ring_plt}
\end{equation}
(c.f. equation 40 of \citealt{Lee22}; note that they had a typo on the power-law index of $a_{\rm ring}$, equivalent to their $R_f$, which we correct here), where $\Sigma_{\rm ring}$ is the solid surface density of the dust ring, and $v_{\rm rms}$ is the velocity dispersion of particles within the ring which we take as $v_{\rm rms} = c_s \sqrt{\alpha / (1 + {\rm St})}$ \citep{Youdin07} assuming the particle random velocities to be coupled to gas turbulence.

As shown in Figure \ref{fig:Sigma_diag}, rings that are at wider orbital separations are more likely to nucleate bound dust clumps, owing to the minimum surface density scaling as $\propto c_s a_{\rm ring}^{-3/2}$ (physically, the tidal shear effect diminishes farther out). Nevertheless, none of the rings studied in this paper are dense enough to nucleate a stable bound clump. Repeating the analysis with using (St/$\alpha$)$_{\rm max}$ uniformly so that the derived $\alpha$ is smaller (and therefore the minimum required $\Sigma_{\rm ring}$ is small; c.f. equation \ref{eq:Sig_ring_plt}) recovers only AS 209 B120 to be able to nucleate a stable bound clump and only at the lowest St (1.21$\times 10^{-3}$) and corresponding $\alpha = 1.33\times 10^{-4}$.

\subsection{Mass growth by pebble accretion} \label{ssec:mass-growth}

Focusing on the most optimistic scenario of a bound clump formation in AS 209 B120 as identified in the previous section, we now evaluate how massive such a clump may grow. First, we evaluate the minimum mass of a bound planetesimal clump following equation 41 of \citet{Lee22}:
\begin{equation}
    M_{\rm pl} > \frac{1}{9}M_\star \left(\frac{c_s^2}{3\pi G\Sigma_{\rm ring}a_{\rm ring}}\right)^3 \left(\frac{\alpha}{1+{\rm St}}\right)^3,
    \label{eq:Mpl_min}
\end{equation}
which evaluates to 2.2$\times 10^{-6} M_\oplus$ and will serve as the minimum initial mass.

Further mass growth of a clump ensues by accreting solids whose accretion rate is generally written as
\begin{equation}
    \dot{M}_{\rm core} = 2\Sigma_{\rm ring}R_{\rm acc}v_{\rm acc}\times {\rm min}(1, R_{\rm acc}/H_{\rm sol})
    \label{eq:Mdot-core}
\end{equation}
where particles that enter within $R_{\rm acc}$ of the protocore at speeds of $v_{\rm acc}$ (measured as the relative speed between the protocore and particles) will be accreted, and $H_{\rm sol} \equiv H\sqrt{\alpha/(\alpha+{\rm St})}$ is the solid disk scale height \citep{Youdin07}.

We consider the theory of pebble accretion to determine $\dot{M}_{\rm core}$ \citep[e.g.,][]{Ormel10}. The core-particle relative velocity $v_{\rm acc}$ is a combination of headwind $v_{\rm hw}$, shear $(3/2)R_{\rm acc}\Omega$, and turbulent random velocity $v_{\rm rms}$ (see, e.g, \citealt{Lin18}, their equation 14; see also \citealt{Ormel18} and \citealt{Rosenthal18}). 
In calculating $v_{\rm hw}$, we account for the deceleration of particles in the pressure trap. Direct numerical simulations find that the peaks of dust density often occur at the radial center of the pressure bump \citep[e.g.,][]{Lee22} and so we evaluate $v_{\rm hw}$ at the maximum size of the clump min($w_d$,$H_{\rm sol}$) away from the center of the pressure bump. The width of the gas pressure bump for AS 209 B120 is precisely determined using gas rotation curves and so we follow equation 2 of \citet{Rosotti20} and derive
\begin{align}
    v_{\rm hw, bump} &= -\left(\frac{c_s^2}{2v_k}\right)\gamma_{\rm bump} \nonumber \\
    &= \left(\frac{c_s^2}{2v_k}\right) {\rm Exp}\left(-\frac{1}{2}\frac{\alpha}{\rm St}\right),
\end{align}
where $w_d < H_{\rm sol}$ for AS 209 B120 and $v_{\rm hw,bump} \sim 980\,{\rm cm\,s^{-1}}$ is $\sim$34\% smaller than $v_{\rm hw}$ in the smooth background disk (c.f. Appendix A of \citealt{Rosotti20}). Because $v_{\rm hw, bump} > v_{\rm rms} \sim 335\,{\rm cm\,s^{-1}}$, we only need to determine whether the accretion is headwind or shear-dominated. Pebble accretion is expected to be headwind rather than shear-dominated when the mass of the accreting core is small (c.f. equation 7.9 and 7.10 of \citealt{Ormel17}; see also equation 24 of \citealt{Lambrechts12}):
\begin{equation}
    M_{\rm core} < \frac{1}{8}\frac{v_{\rm hw, bump}^3}{G\Omega}{\rm St}^{-1}.
    \label{eq:Mcore-trans}
\end{equation}
This maximum mass for headwind-dominated accretion is $\sim$1.8$M_\oplus$ for AS 209 B120 and we will show that a dust clump would never exceed this mass. We therefore safely assume $v_{\rm acc} \sim v_{\rm hw, bump}$ at all times.

If pebble accretion were to occur, the particle stopping time must be shorter than the encounter time (${\rm St}/\Omega < R_{\rm acc}/v_{\rm acc}$; the settling regime, see \citealt{Ormel10}). We first assume this to be true to compute $R_{\rm acc}$ and verify a posteriori that the condition is indeed satisfied. Under the settling condition,
\begin{equation}
    v_{\rm acc} = 4\frac{GM_{\rm core}}{R_{\rm acc}^2}\frac{\rm St}{\Omega}
    \label{eq:vacc-settl}
\end{equation}
so that the particles that accrete reach a terminal velocity during the encounter with the core. Substituting $v_{\rm acc} = v_{\rm hw,bump}$ to solve for $R_{\rm acc}$, the settling condition is satisfied when
\begin{equation}
    M_{\rm core} > \frac{1}{4}\frac{v_{\rm hw,bump}^3}{G\Omega}{\rm St}
    \label{eq:Mcore-settl}
\end{equation}
which is larger by a factor of 2 compared to equation 7.11 of \citet{Ormel17} because of different assumptions of order unity factors in the settling condition and in equation \ref{eq:vacc-settl}. For AS 209 B120, this minimum mass for pebble accretion is $\sim$5.23$\times 10^{-6}M_\oplus$ which is larger than the minimum initial clump mass (equation \ref{eq:Mpl_min}) so we use equation \ref{eq:Mcore-settl} as the initial core mass for pebble accretion.

In equation \ref{eq:Mdot-core}, the accretion proceeds in a three-dimensional fashion when $R_{\rm acc} < H_{\rm sol}$:
\begin{equation}
    M_{\rm core} < \frac{1}{4}\frac{v_{\rm hw,bump}\Omega H^2}{G{\rm St}}\left(\frac{\alpha}{\alpha+{\rm St}}\right).
    \label{eq:Mcore-3d}
\end{equation}
Headwind-dominated accretion will always be in three-dimensional regime if the maximum mass in equation \ref{eq:Mcore-trans} is smaller than that in equation \ref{eq:Mcore-3d} which translates to
\begin{equation}
    \frac{1}{2}\left(\frac{v_{\rm hw,bump}}{c_s}\right)^2 < \frac{\alpha}{\alpha+{\rm St}},
\end{equation}
which is satisfied for AS 209 B120.

The accretion rate then follows
\begin{align}
    \dot{M}_{\rm core} &= \frac{8\Sigma_{\rm ring}G M_{\rm core}{\rm St}}{c_s}\left(\frac{\alpha+{\rm St}}{\alpha}\right)^{1/2} \nonumber \\
    &=\left(\frac{2}{\pi}\right)^{3/2}\left(\frac{v_k}{c_s}\right)^2\frac{M_{\rm ring}(t)}{M_\star}M_{\rm core}\Omega\frac{H}{w_d}{\rm St}\left(\frac{\alpha+{\rm St}}{\alpha}\right)^{1/2},
    \label{eq:Mdot-core-settl}
\end{align}
and
\begin{equation}
    M_{\rm core}(t) = M_{\rm pl,0}\,{\rm Exp}\left[F(M_\star,a_{\rm ring},\alpha,{\rm St})\int^t_0 M_{\rm ring}(t')dt'\right]
    \label{eq:Mcore-pebbl-growth}
\end{equation}
where $M_{\rm pl,0}$ is set by either equation \ref{eq:Mpl_min} or equation \ref{eq:Mcore-settl} whichever is larger, $F(M_\star, a_{\rm ring}, \alpha, {\rm St})$ encapsulates all the other time-invariant quantities in equation \ref{eq:Mdot-core-settl}, and $M_{\rm ring}(t')$ follows from equation \ref{eq:Mring}.

For constant particle size, taking $a_0$ from equation \ref{eq:a0_fixs}:
\begin{align}
    &\int^t_0 M_{\rm ring}(t')dt' = M_0\epsilon_{\rm trap} \times \nonumber \\
    &\left[-\frac{(A+De^{Ct'})^{1+\beta}\,_2F_1(1,1+\beta,2+\beta,1+\frac{D}{A}e^{Ct'})}{AC(1+\beta)} - a_{\rm ring}^\beta t'\right]^t_0
    \label{eq:Mring-integ-fixs}
\end{align}
where
\begin{equation}
    A \equiv -\frac{3}{2}\frac{\alpha}{|\gamma_{\rm bump}|{\rm St}_r}a_{\rm ring},
\end{equation}
\begin{equation}
    D \equiv a_{\rm ring} - A,
\end{equation}
\begin{equation}
    C \equiv \frac{c_{s,r}^2}{v_{k,r}}|\gamma_{\rm bump}|{\rm St_r}a_{\rm ring}^{-1},
\end{equation}
and $_2F_1$ is the hypergeometric function.

\begin{figure}
    \centering
    \includegraphics[width=0.5\textwidth]{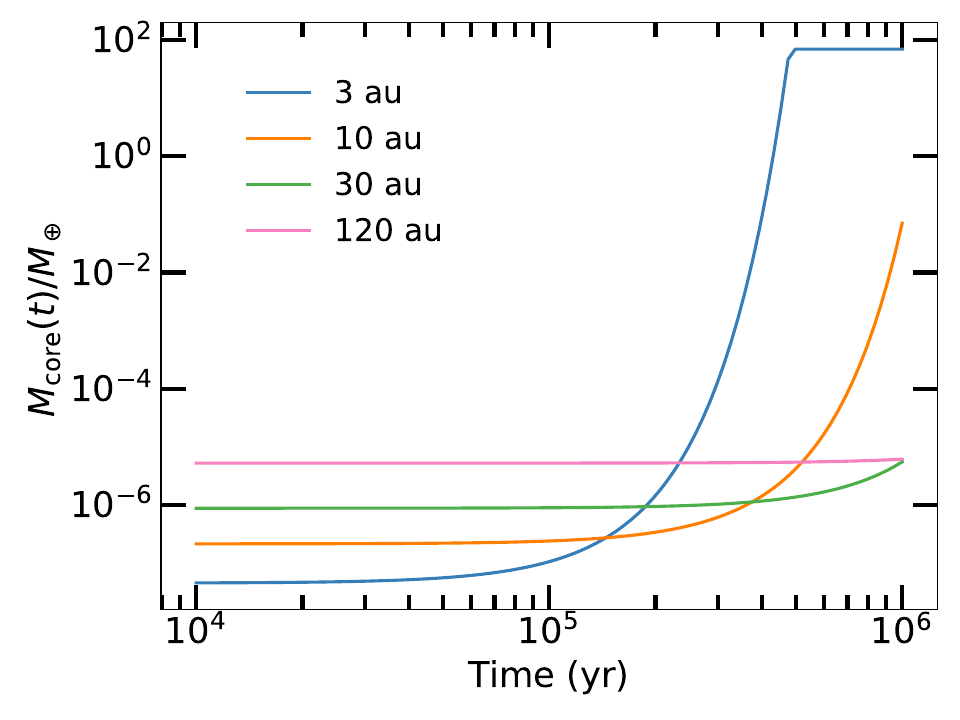}
    \caption{Mass growth of a dust clump within AS 209 B120 (pink line). Shown in other lines are the rescaled mass growth if the ring were to be placed at closer-in distances with the same $M_{\rm ring}$, $w_d/a_{\rm ring}$, St, and $\alpha$. Planet formation is prohibitively slow at the measured location of the ring but accelerates dramatically inside 10 au, potentially engulfing all the ring masses if placed at 3 au.}
    \label{fig:Mcore}
\end{figure}

Figure \ref{fig:Mcore} demonstrates minimal core growth within AS 209 B120 (pink line). The low St, $\alpha$, and the wide orbital separation all conspire to render core growth prohibitively sluggish. We therefore conclude that the observed rings are stable against clump formation or even if it nucleates dust clumps, they are expected to be destroyed (primarily by tidal shear) and even in a marginal case where a bound clump can form, such clumps would not have enough time to grow in mass, ensuring the survival of the rings as they are observed.

Our finding of minimal planet formation in dust rings differs from that of \citet{Lau22} and \citet{Jiang23} who argue for rapid planet formation in dust rings \citep[but see][]{Morbidelli20}. First, both studies assume stable planetesimal formation in rings: \citet{Lau22} use a prescribed activation function that is independent of local turbulence while \citet{Jiang23} set up their rings so that the local dust-to-gas ratio is already one. We find that such prescribed conditions are not realized in observed rings nor would the clumps be stable if formed. Second, \citet{Lau22} adopt a distribution of St that samples high values $\sim$0.1 which dominates the dust mass (see their Figure 3) and would dramatically accelerate the rate of pebble accretion. We find that such high St are not likely in the rings we study. \citet{Jiang23} on the other hand explore low St $\sim$10$^{-3}$ with default $\alpha = 10^{-3}$ so at face value, we would expect planet growth to be even more impeded under their condition. The reduction by higher $\alpha$ is dwarfed by their choosing dense $\Sigma_{\rm ring}$ which would exponentially accelerate the rate of pebble accretion (see equation \ref{eq:Mdot-core-settl}). From their equation 34, the dust surface density of AS 209 B120 would correspond to $\sim$1 g cm$^{-2}$ which is about an order of magnitude larger than the observed value.

\subsection{Implication on planet formation} \label{ssec:pl-form}

One caveat to our conclusion above is that we ignore the possibility of creating multiple clumps. Without direct numerical simulation, it is not clear how many bound clumps may form. Nevertheless, we can estimate the number using the maximum size of the clump as the dust scale height:
\begin{align}
    N_{\rm cl} &\sim \frac{2\pi a_{\rm ring}}{H_{\rm sol}} \nonumber \\
    &\sim \frac{2\pi a_{\rm ring}}{H}\sqrt{\frac{\alpha+{\rm St}}{\alpha}}
\end{align}
which is $\sim$170 for AS 209 B120. Unless all the clumps are initially about half an Earth mass---which is more massive than typically quoted masses of initial planetesimals \citep[e.g.,][]{Cuzzi10,Gerbig23}---the total amount of mass locked within these clumps would be negligible compared to the total ring dust mass, also consistent with the existence of rings.

By contrast, if we compute the same analysis with the ring placed at closer-in distances, planet formation becomes much more efficient, so much so that the entire ring can be engulfed into a planet inside 10 au (see Figure \ref{fig:Mcore}). In this experiment, we keep $M_{\rm ring}$, $w_d/a_{\rm ring}$, St, and $\alpha$ fixed and scale all other quantities with $a$, including $\phi \propto a^{2/7}$ \citep{Chiang97} so that the accelerated growth at shorter $a$ is due to the shorter dynamical time \citep[see also][]{Chambers21}.\footnote{At these shorter distances, pebble accretion becomes shear-dominated but remain in three-dimensional regime so that $\dot{M}_{\rm core}$ and $M_{\rm core}(t)$ still follow equations \ref{eq:Mdot-core-settl} and \ref{eq:Mcore-pebbl-growth}.} 

This strong sensitivity to the radial location of the ring has an interesting implication on the connection between disk and exoplanet observations. In a recent review, \citet{Bae23-PPVII} find that disks with substructures tend to be of high disk mass (see their Figure 3, panel a), whereby the mass is obtained from the measured mm-flux assuming optically thin limit. Given that larger disks tend to also be luminous \citep[e.g.,][]{Tripathi17, Zormpas22}, we may deduce that disks with substructures tend to be large. As cautioned by \citet{Bae23-PPVII}, the observed lack of rings in lighter (smaller) disks may be an observational bias as such disks lack the required high resolution data, but the results presented in this paper would suggest that this may be physical. It may be that even if such rings existed, they may quickly have transformed into planetary mass objects and therefore no longer observable. Such a prospect is also consistent with current exoplanet demographics. Direct imaging surveys report a decline in the population of gas giants beyond 10 au \citep[e.g.,][]{Nielsen19}, which is also consistent with the long baseline radial velocity surveys \citep[e.g.,][]{Fulton21} that show a peak in the giant occurrence rate at 1--10 au. Our calculation demonstrates that the initial growth of planets is difficult beyond $\sim$10 au even within dust rings where we see concentration of mass. By contrast, inside $\sim$10 au, formation of massive cores becomes precipitously more likely.

Our calculation further demonstrates that at $\sim$10 au and beyond, rocky bodies can at best be a Mars mass individually (see orange curve in Figure \ref{fig:Mcore}). While planets are often invoked as the source of the rings \citep[e.g.,][]{Teague21}, conclusive evidence remains elusive for many of the observed ringed disks \citep{Speedie22}. Even if the first generation planets did indeed create some of these rings, such rings are expected to be stirred up \citep[e.g.,][]{Bi21,Bi23}, hindering both planetesimal formation and pebble accretion and therefore the generation of second population of planets. As long as the rings sampled in this study are a good representation of {\it typical} planet-forming disks, our results would predict that, like the gas giants, we would see a similar decline in the population of smaller Neptunes and super-Earths beyond $\sim$10 au, to be confirmed by the Roman Space Telescope.

\section{Conclusion} \label{sec:concl}

By coupling the measured properties of dust rings in nearby bright T Tauri disks with Class 0/I disk using dust radial equation of motion, we have separately determined local St and $\alpha$ in ringed disks at large orbital distances. We find that both parameters are low ($10^{-4} \lesssim {\rm St} \lesssim 10^{-2}$; $10^{-5} \lesssim \alpha \lesssim 10^{-3}$) consistent with inviscid disks where Type I migration would be prematurely halted. We further find that given such low St, the estimated local $\Sigma_g$ is high so that the dust-to-gas ratio is too small to trigger streaming instability for the corresponding $\alpha$. Even if initial clumping may be possible by other dust-driven instabilities, we find that the rings are not dense enough to nucleate clumps that are stable against both turbulent diffusion and tidal shear. In one marginal exception, at the lowest St and $\alpha$, AS 209 B120 may be able to create bound clumps but at such wide orbital distances, combined with low St and $\alpha$, core growth by pebble accretion is minute. Our analysis is consistent with the fact that the rings are observable over the system age.

An interesting implication of our analysis is that beyond $\sim$10 au, we may see a decline in the planet population even for Neptune-class objects, similar to the already observed decline in wide-orbit gas giants, insofar as the ringed disks being typical planet-forming disks. Such a prediction should be testable by space-based microlensing studies such as the Nancy Grace Roman Space Telescope.

\vspace{0.5cm}
The anonymous referee provided an encouraging and insightful report that helped improve the manuscript. I thank Wlad Lyra and Wenrui Xu for their questions at the Extreme Solar Systems V which motivated some of the supporting calculations and Ruobing Dong for useful discussions. E.J.L. gratefully acknowledges support by NSERC, by FRQNT, by the Trottier Space Institute, and by the William Dawson Scholarship from McGill University.

\appendix
\counterwithin{figure}{section}

\section{Limits on Stokes number for each system}

\begin{figure}
    \gridline{\fig{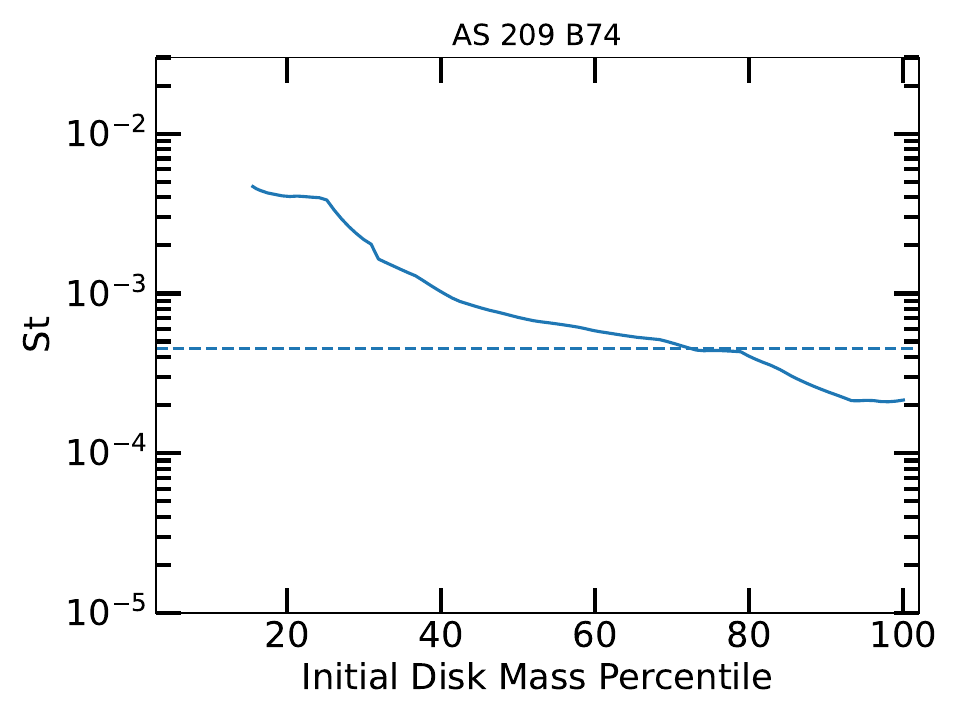}{0.4\textwidth}{}
            \fig{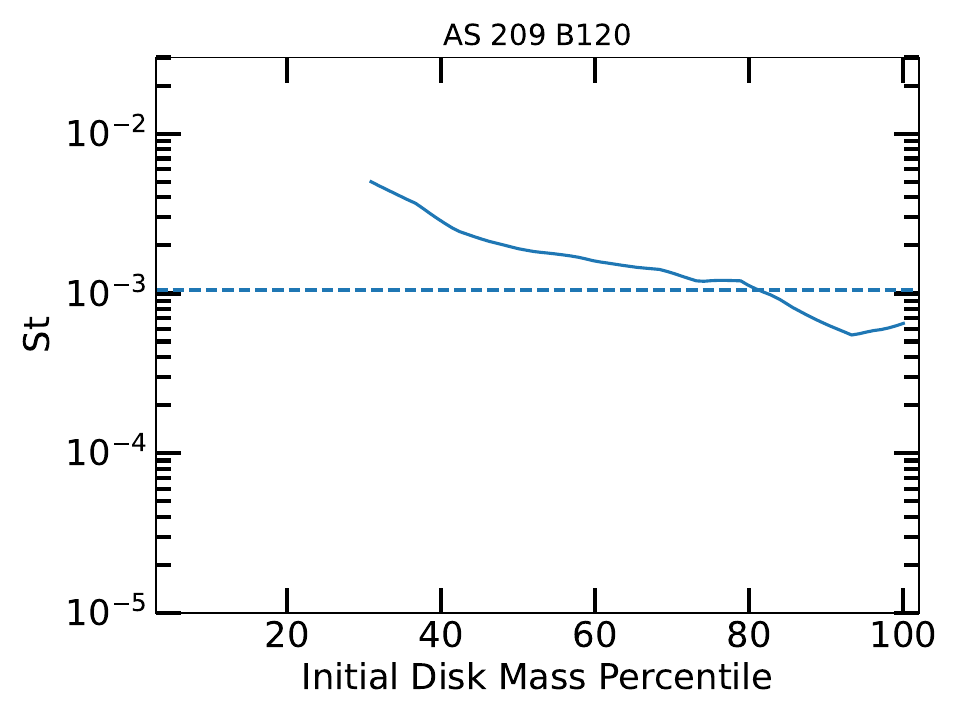}{0.4\textwidth}{}}
            \vspace{-0.7cm}
    \gridline{\fig{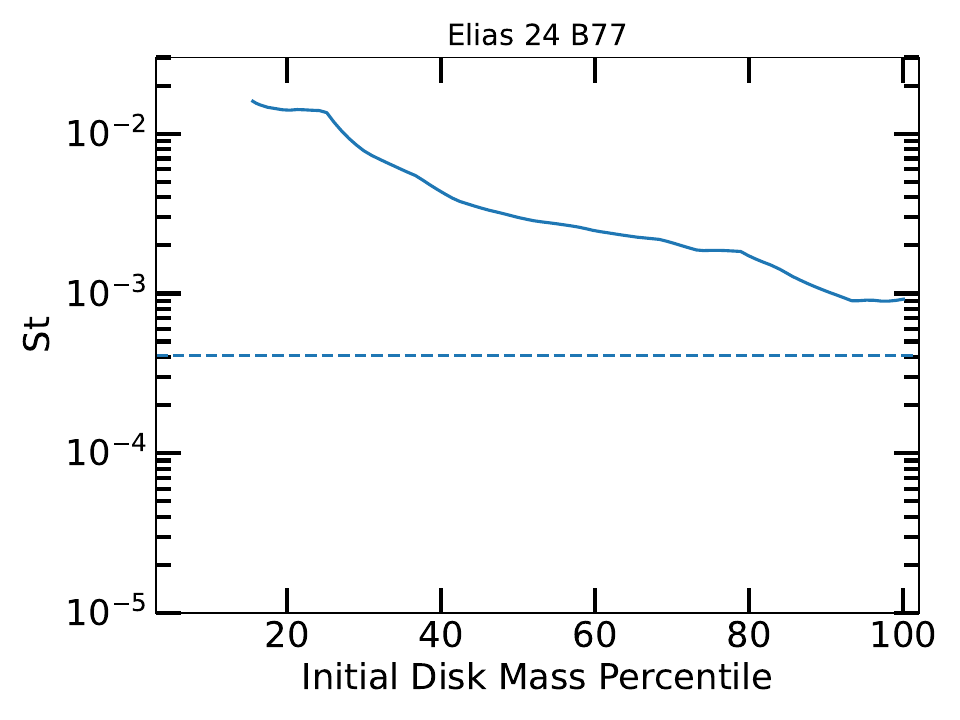}{0.4\textwidth}{}
            \fig{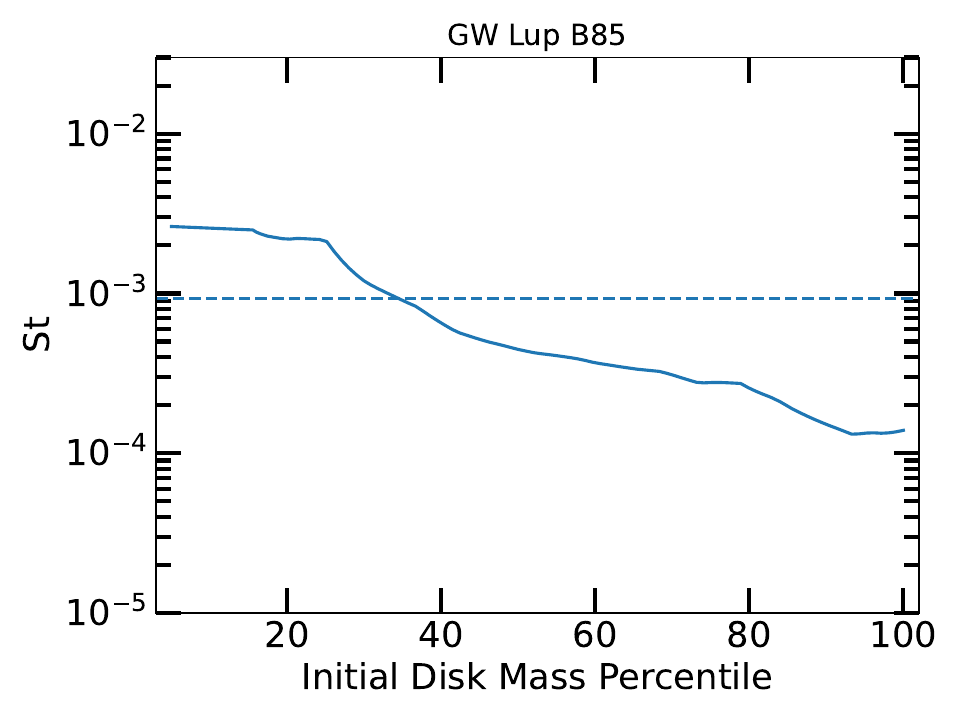}{0.4\textwidth}{}}
            \vspace{-0.7cm}
    \gridline{\fig{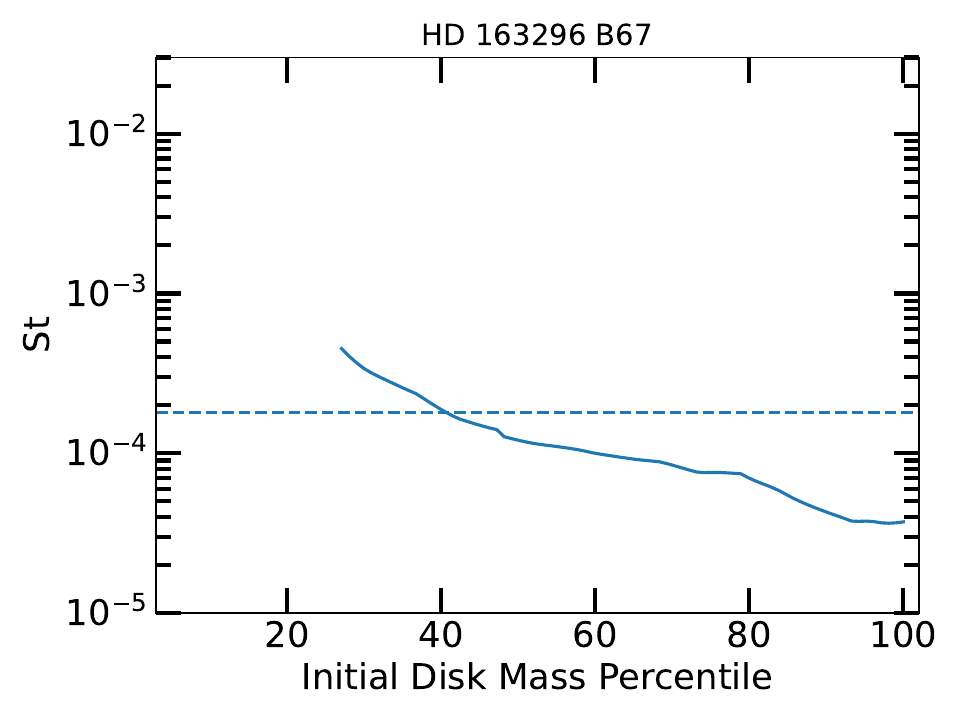}{0.4\textwidth}{}
            \fig{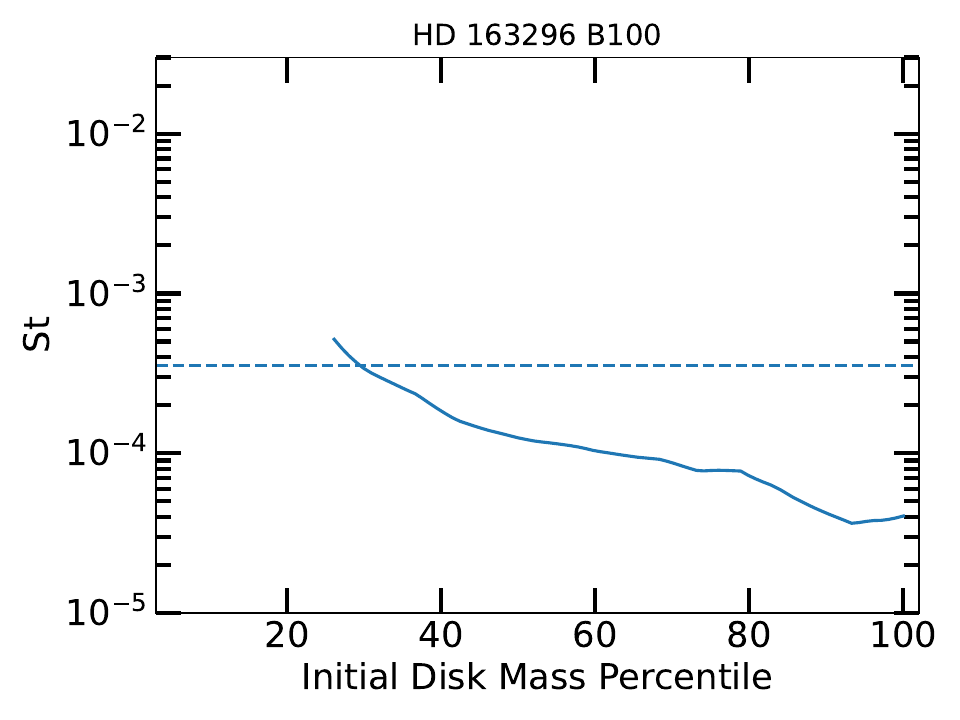}{0.4\textwidth}{}}
            \vspace{-0.7cm}
    \gridline{\fig{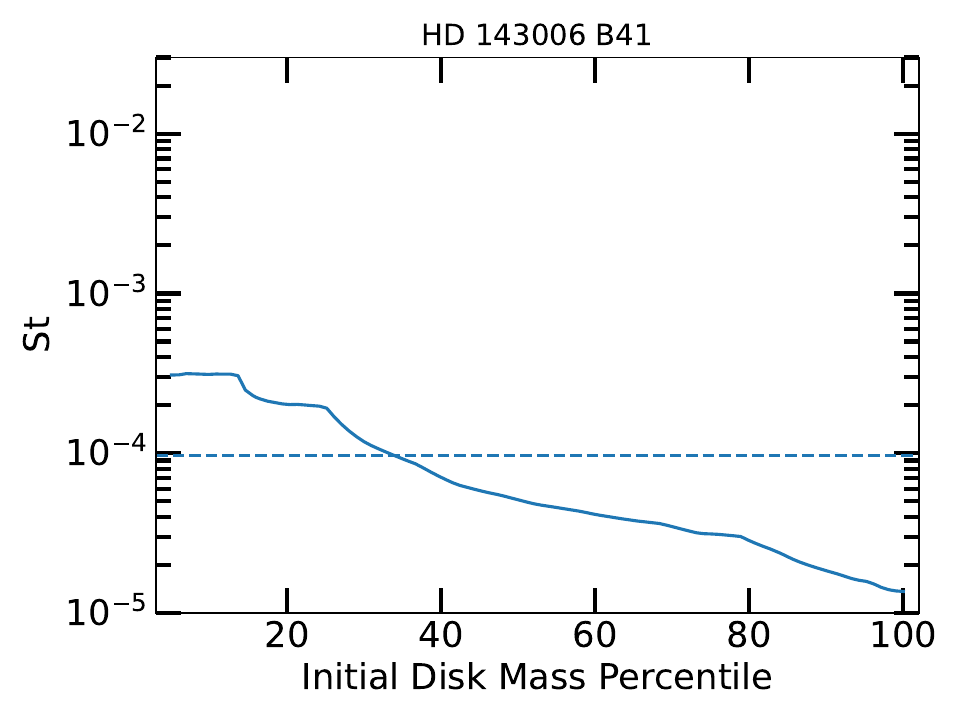}{0.4\textwidth}{}
            \fig{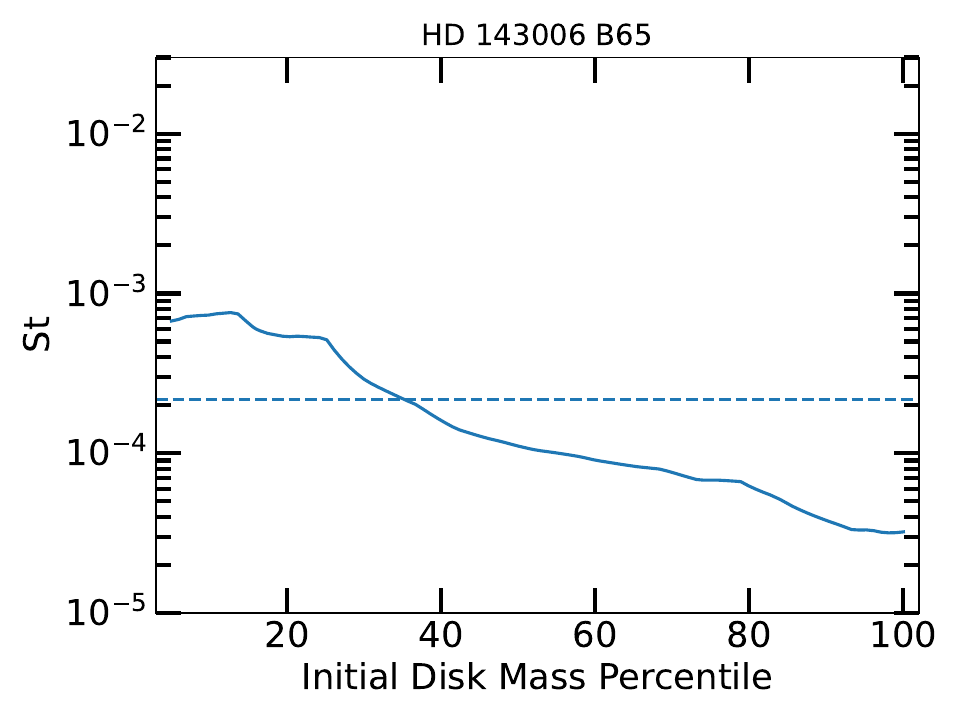}{0.4\textwidth}{}}
            \vspace{-0.7cm}
    \caption{Derived St as a function of initial disk dust mass percentiles. For each system, the minimum St required for local gravitational stability is drawn in horizontal dashed line.}
    \label{fig:St_Md0_ring}
\end{figure}

We illustrate in Figure \ref{fig:St_Md0_ring} the range of allowable St for each dust rings over 5th and 100th percentile of initial disk mass as shown in Figure \ref{fig:Mdust_disk}.

\bibliography{ringed-disk}{}
\bibliographystyle{aasjournal}

\end{document}